\documentstyle[prl,aps,multicol,epsfig]{revtex}
\begin{document}
\draft
\widetext

\title{Theory for the excitation spectrum of High-T$_c$
superconductors :\\
quasiparticle dispersion and  shadows of the Fermi
surface}
\author{ M. Langer, J. Schmalian, S. Grabowski,
and K. H. Bennemann}
\address{ Institut f\"ur Theoretische Physik,
 Freie Universit\"at Berlin, Arnimallee 14, 14195
Berlin , Germany}

\date{15 may 1995}
\maketitle

\widetext
\begin{abstract}
\leftskip 54.8pt
\rightskip 54.8pt
Using a new   method for the solution of the
FLEX-equations,
which allows the determination of the   self
energy $\Sigma_{\bf k}(\omega)$
of the $2D$  Hubbard model  on the real
frequency
axis, we calculate  the doping dependence of
the quasi-particle excitations
  of   High-T$_c$ superconductors.
We obtain   new results  for the  shadows of
the Fermi surface,
their dependence on the deformation  of the
quasi particle dispersion,
an anomalous $\omega$-dependence of
${\rm Im}\Sigma_{\bf k}(\omega)$
and a related violation of the Luttinger
theorem.
 This  sheds new  light on the influence of
 short range magnetic
order on the low energy excitations and
its significance for
photoemission experiments.
 \end{abstract}


\begin{multicols}{2}
%%%%%%%%%%%%%%%%%%%%

\narrowtext

%\widetext
%\newpage

%
Despite an enormous progress~\cite{D94},  the electronic
excitation spectrum of the strongly
correlated High-T$_c$ superconductors is still far
from being  understood.
 Recent angular resolved photoemission
measurements~\cite{LVP92,DSK93,AOS94,GCA94,WSM95}
for different doping concentrations
 demonstrate that pronounced deformations of
the quasiparticle dispersion occur.
The opening of a spin density gap and the
variation in spectral weight as
 function of ${\bf k}$~\cite{WSM95} reflect
the   strong influence of the antiferromagnetic
correlations on the  low energy excitations.
In particular, the interpretation of the
shadows  of the Fermi surface (FS)
observed    by Aebi {\em et al.}~\cite{AOS94}
in terms of  antiferromagnetic
correlations is    under current debate~\cite{C95}.
Therefore, the determination of the  elementary
excitations  with high
energy and momentum resolution is of extreme
importance.

In this Letter, we calculate the quasiparticle excitation
spectrum of the one band Hubbard
Hamiltonian using a new numerical method for the
self consistent summation of all  bubble and
ladder diagrams~\cite{BS89} (fluctuation exchange
approximation) on the real
frequency axis, and compare our results with recent
photoemission experiments.
Due to the high degree of numerical stability of this
method, we present interesting new results for
 a strong   deformation of the quasiparticle dispersion
at the  wave vector ${\bf k}=(\pi,0)$
upon doping, the occurrence of  satellite peaks and
shadows of the Fermi surface in the
paramagnetic state and an unusual momentum and
frequency dependence of the electronic
self energy.
All this sheds new light on the occurrence of FS-shadows w
ithout long range antiferromagnetic
order.

The   fluctuation exchange (FLEX)
approximation~\cite{BS89,BSW89,SH91,DT95}
is  conserving   in
the sense of Kadanoff and Baym~\cite{BK61}.
It is a theoretical approach complementary to the exact
diagonalization studies~\cite{DNB94} or   quantum
Monte Carlo simulations~\cite{BSW94,PHvdL95},
which gave deep insight
into the origin of the quasiparticles of strongly correlated
systems, but   are limited to rather
small  systems.
The self energy   in the FLEX  approximation  of the
paramagnetic phase
is given by the momentum and Matsubara-frequency sum
 $\Sigma(k)=\frac{T}{ N} \sum_{  k'} V ( k-k')  G ( k')$,
 where  $k=({\bf k},i\omega_n)$.  The effective    i
nteraction $V (q)$   results from the
summation of bubble and ladder diagrams  and can
be expressed in terms
of the  particle-hole bubble
  $\chi(q)  = -\frac{T}{ N} \sum_{   k }
 G (k+q ) G ( k ) $.~\cite{BS89}
$G  (k )$ is the fully renormalized
Greens function,  $ \omega_n $   a fermionic
Matsubara frequency,
${\bf k}$ the crystal momentum,  $N$ the number
of lattice sites, and $T$ the temperature.

Because of the straightforward applicability
of the numerically
very effective fast Fourier transformation (FFT)
for the evaluation of the above convolutions,
most  of the  solutions of the FLEX equations
are performed on the imaginary
frequency axis.
Unfortunately,  from the imaginary axis,  reliable
information about the
real frequency excitations can only be obtained
using rather ill defined  numerical
procedures.
Based on the contour integral technique, we present
in the following a numerically stable calculation
of the electronic self energy on the real axis using
the FFT.
Note that recently Dahm {\em et al.}~\cite{DT95}
also solved the FLEX equations on the real axis.
However, they performed   the  direct calculation
of  the rather time consuming  frequency integrals.

 In the following, we first  apply our  approach to the
one band Hubbard model
in the paramagnetic state.
The bare  dispersion is given by
$\varepsilon^o_{\bf k}=-2t_o(\cos(k_x)+\cos(k_y))-\mu$,
with nearest neighbor hopping element $t_o=0.25 \, {\rm eV}$,
and  chemical potential $\mu$.
The local Coulomb repulsion  is given by $U =4t_o$.
The extension to multi band Hamiltonians and to
superconducting or magnetic  phases is straightforward.~\cite{GSL95}
 We solve the set of equations   for  complex frequencies
 $z=\omega+i\gamma$ with small but finite imaginary
part $\gamma < \pi T$.
Then, the analytical continuation to the real
axis $\gamma \rightarrow 0^+$ is  numerically well defined.

The above  momentum summations become simple products
in Wannier space.
Furthermore, the frequency summations  are evaluated using
the  contour integral
technique~\cite{AGD65} , where we place the four horizontal
lines of the contour  by a finite amount
$\gamma $ away from the poles of the Greens function.
 Considering the  particle hole bubble  $\chi_{{\bf i}}(z)$
at lattice site ${\bf i}$ and for frequencies $z=\omega +i\gamma$, we
perform the Kramers Kronig transformation for the Greens function
and express the resulting
energy denominators  using Laplace transformation~\cite{CS90}.
  Now, the occurring frequency integrals decouple and we obtain
$\chi_{{\bf i}}( z) =\int_0^\infty dt \,  \chi_{{\bf i}}(t)\,
e^{i zt}$ with
\begin{eqnarray}
\chi_{{\bf i}}(t)&=&-i(2 \pi)^2 e^{-\gamma t}\left( \varrho_{{\bf i}}(t)
[{\cal A}^\ast_{-{\bf i}}(t)  +
    {\cal A}_{-{\bf i}}(-t) e^{2\gamma t} ] \right. \nonumber \\
& & \left. -\varrho^\ast_{-{\bf i}}(t)  [{\cal A}_{{\bf i}}(t)+
{\cal A}^\ast_{{\bf i}}(-t)  e^{2\gamma t} ]\right)\, .
\label{chi_t}
\end{eqnarray}
Here, ${\cal A}_{{\bf i}}(t)$ is defined by
\begin{equation}
{\cal A}_{{\bf i}}(t)=\frac{i}{2 \pi} \int_{-\infty}^{\infty}
\frac{d \epsilon}{2 \pi}  f^\ast(\epsilon +i \gamma)
G^\ast_{{\bf i}}(\epsilon +i \gamma)   \, e^{-i \epsilon t}\, ,
\label{A_t}
\end{equation}
and $  \varrho_{{\bf i}}(t)$  is the Fourier transform of the
spectral density $-\frac{1}{\pi} {\rm Im}
G_{{\bf i}}(\epsilon +i \gamma)$.
 Note that we are using the complex Fermi function $f(z)$,
because we solve the FLEX equations
  consistently for finite $\gamma$. This is essential for the
fine structure at small energies.
Thus $\chi_{\bf k}( z)$ and finally $V_{\bf k}(z)$ can be
evaluated by a $D+1$
 dimensional   FFT of $\chi_{{\bf i}}(t)$.
The procedure  for the frequency summation   of  the
self energy is similar.
Within the Wannier space, it follows for the Laplace
transform of  the  self energy:
\begin{eqnarray}
\Sigma_{{\bf i}}(t)&=& -i 2 \pi  T V_{{\bf i}}(0)\,
\varrho_{{\bf i}}(t)
+i(2 \pi)^2 e^{-\gamma t}\left( v_{{\bf i}}(t) [{\cal A}_{{\bf i}}(t)
+ \right.   \nonumber \\
 & &    \left.    {\cal A}^\ast_{{\bf i}}(-t)   e^{2\gamma t} ]
  -\varrho_{{\bf i}}(t) [{\cal B}^\ast _{{\bf i}}(t)  +
{\cal B}_{{\bf i}}(-t) e^{2\gamma t} ]\right)\, .
\label{sigma_t}
\end{eqnarray}
$v_{{\bf i}}(t)$ is the Fourier transform of $- \frac{1}{\pi}
{\rm Im } V_{{\bf i}}( z)$
and ${\cal B}_{{\bf i}}(t)$ can be obtained from Eq.~\ref{A_t}
by replacing $f^\ast( z)
G^\ast_{{\bf i}}( z) $ by $n^\ast( z)  V^\ast_{{\bf i}}( z) $.
 $n(z)$ is the complex Bose function, and the first term
results from the  coincidence
of the fermionic Matsubara frequencies
$\omega_n=\omega_{n'}$, which cannot be
included in the  standard contour.
Thus, the self energy $\Sigma_{\bf k}(\omega +i\gamma)$
results via $D+1$ dimensional
  FFT~\cite{asympt} from  $\Sigma_{{\bf i}}(t)$ and the limit
$\gamma \rightarrow 0^+$ can be performed.

The  advantage of our method compared to the
first real frequency calculation
recently published by Dahm {\em et al.}~\cite{DT95}
is the applicability of the FFT and the consistent
solution
for  a small but finite imaginary part $\gamma$,
which lead to an accuracy of the convergence
of the spectral density per momentum and
frequency point  of $10^{-5}$.
Only this accuracy enables us to obtain the  new
physical results presented  here\cite{details}.

\begin{figure}
\vskip -1cm
\centerline{\epsfig{file=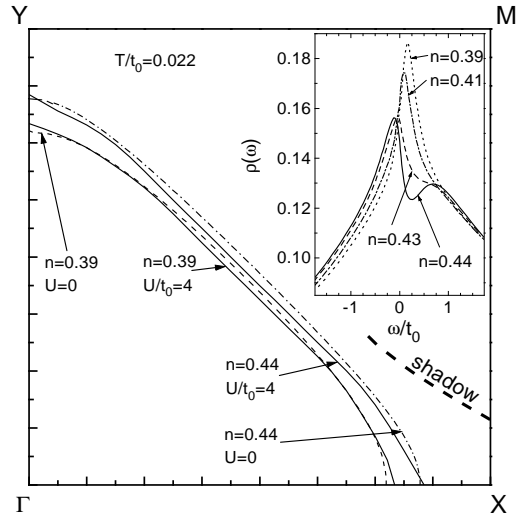,width=9cm,height=12cm}}
\vskip -3cm
\caption{ Fermi surface in the quarter of the
Brioulline zone with $k_x, k_y > 0$ (solid
line) in comparison with the Fermi surface of
the uncorrelated system
for different doping concentrations. Note the
violation of the Luttinger theorem and the occurrence
of FS-shadows for smaller doping ($n=0.44$).
The inset shows the local density  of states for various
doping values. Note, the occurrence of a pseudogap
near the half filled case.}
 \label{fig1}
 \end{figure}

In the inset of Fig.~\ref{fig1}, we show results for
the local (momentum averaged)
density of states (DOS)  for various doping values.
 We find for larger doping a rigid band like behavior, but
for smaller doping   a  pseudogap occurs as a precursor
of the Mott-Hubbard splitting.
{}From the asymmetry with respect to the pseudogap,
a transfer of spectral weight from high energy to low
energy scales~\cite{EMS91} can be seen.

In  Fig.~\ref{fig1}, results are shown for the FS for
different  band fillings
in comparison with that for $U=0$.
The FS is obtained from the ${\bf k}$-points, where
 $\varepsilon^o_{\bf k}+{\rm Re}\Sigma_{\bf k}(0)=0$.
The changes of the FS are rather small, but a tendency
to a stronger  nesting topology   is clearly visible.
 Such a deformation of the FS is related to a dramatic
deformation
 of the quasi particle dispersion.
For larger doping, the Luttinger theorem~\cite{L61} is
fulfilled.
However, for smaller doping  the volume of the FS
decreases compared to the $U=0$ case.
This violation of the Luttinger theorem results from a
transfer of particles to additional
shadow states (schematically shown in Fig.~\ref{fig1}),
as will be discussed below.

\begin{figure}
\centerline{ \epsfig{file=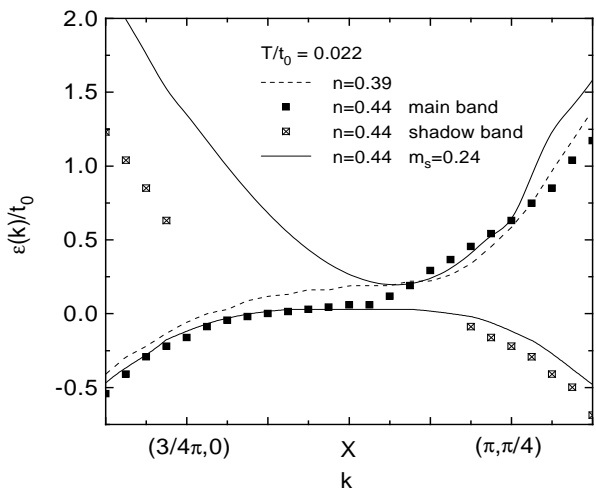,width=9.5cm,height=12cm}}
\vskip -3cm
\caption{ Quasiparticle dispersion  in the neighborhood
of the $X$-point for different doping
  concentrations.
 Note that for smaller doping ($n=0.44$) two bands exist
near the Fermi energy  in the
paramagnetic state. The dispersion of the main band
(solid squares) resembles that of the
highly doped system (dashed line).
Considering both, the main band and the shadow band
(open-crossed squares) one recognizes
the evolution towards the two branches of the dispersion
in an antiferromagnetic
 background (solid lines).}
\label{fig2}
 \end{figure}

In   Fig.~\ref{fig2}, we show our results for the quasi
particle dispersion obtained from the maxima of
the spectral density.
Since we expect the FLEX   to be a good approximation
 for the  low energy excitations,
we focus   on  the   dispersion  near the Fermi energy
and in particular
in the neighborhood of the $X$-point.
For larger doping (dashed line), the saddle point,
responsible for the
van Hove singularity, is visible.
However, for smaller doping (solid and open-crossed
squares)   pronounced deformations
of the dispersion occur.
 The dispersion of the states with high spectral weight
(solid squares) is flattened
and suddenly repelled from the Fermi energy.
This might be a precursor of the  large spin density gap
 at the $X$-point,  found in recent
photoemission   experiments by Wells {\em et al.}~\cite{WSM95}
for undoped cuprates.
Furthermore, new quasiparticle states   with low spectral
weight (open-crossed   squares)
 occur.
These states, which are   for larger doping only visible
far away from the Fermi level (not shown),
remain stable for low excitation energies and
 build up the shadows of the
Fermi surface~\cite{AOS94}, which one usually
would expect only for systems with
long range antiferromagnetic order.
In order to check this interpretation, we
performed for the first time
the spin resolved FLEX approximation  within a given
antiferromagnetic  background (solid line).
We fixed the staggered moment $m_{\rm s}$  by its
Hartree Fock
value, because for the doping values under consideration
no long range   order occurs.
 The dispersion in this  antiferromagnetic background
is a clear  continuation  of the evolution of the paramagnetic
state with decreasing doping.
  Obviously, a clear precursor of the     antiferromagnetic state
occurs already in the paramagnetic phase   without broken symmetry.
\vskip -1cm

\begin{figure}
\centerline{\epsfig{file=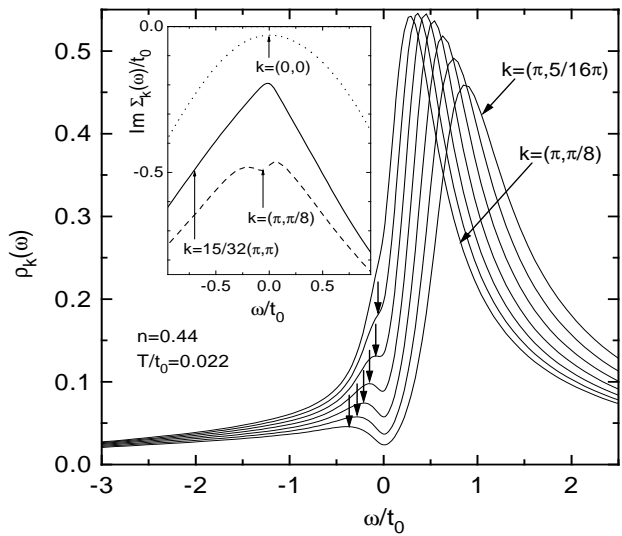,width=9cm,height=11cm}}
\vskip -2cm
\caption{  Spectral density for ${\bf k}$-points between
 the $X$ and $M-$point, i.e.
${\bf k}=(\pi,\frac{l}{32} \pi)$ with $l=4 \cdots 10$.
The main peak reflects the dispersion of the bands
similar to the case for $U=0$.
The satellites, indicated by the arrows, build up
the shadows of the Fermi surface.
The inset shows the
imaginary part of the self energy for different momenta.}
\label{fig3}
 \end{figure}

In Fig.~\ref{fig3}, we show   the occurrence of shadows
of the FS by
plotting the spectral density $\varrho_{\bf k}(\omega)$
for ${\bf k}$-points between the $X$   and
$M$-point.
Besides the main peak, not crossing the FS, satellite
peaks near the Fermi energy can clearly be
observed.
These satellite peaks are the new quasi particle states
of Fig.~\ref{fig2} (open-crossed  squares),
which build up the shadows of the FS.
For  the wave vector ${\bf k}=(\pi, \pi/8)$, which is shifted
by $(\pi,\pi)$ with respect to the main FS,
the shadow band crosses the Fermi level.
The intensity of the shadow states (ca. $10\%$ of the
main peak)  agrees with the experimental
 observation~\cite{AOS94}.
The possible occurrence of such shadow states for
a system without long range antiferromagnetic
order was originally proposed by Kampf and
Schrieffer~\cite{KS90}.
Using a phenomenological ansatz for the spin
susceptibility, they argued
that for sufficient long ranged antiferromagnetic
correlations a coupling of states ${\bf k}$ and
${\bf k}'$ with $|{\bf k}'-{\bf k}-{\bf Q}| < \xi^{-1}$
might lead to distinct shadow states.
{}From their calculation one can estimate the magnetic
correlation length $\xi$ to be of the order of
 $20$ lattice spacings, to obtain observable satellite
 peaks.
This is much larger than the correlation length observed i
n neutron scattering   experiments~\cite{TGS92},
which is only a few lattice spacings.
\vskip -1cm
Therefore, it was argued in Ref.~\cite{C95} that there
 cannot be a well defined kinematics of the
antiferromagnetic spin excitations that lead to the
 observations of Aebi {\em et al.}~\cite{AOS94}.
However, in our calculation the correlation length,
obtained from the ${\bf k}$-dependence of
$V_{\bf k}(\omega)$, is $2.5$ lattice spacings, in
agreement with those experiments.
Furthermore, we find that for larger doping ($n=0.4$)
the shadow states at low energies vanish,
although  the correlation length only  decreases slightly.
In order to resolve a shadow peak at ${\bf k}' = {\bf k} + {\bf Q}$,
the exitation energies of the
 states near ${\bf k}$ should be within an energy range
smaller than the
inverse  lifetime of the state ${\bf k} + {\bf Q}$, i.e.
$ |\varepsilon_{\bf k} - \varepsilon_{{\bf k}+\xi^{-1}}|
< |{\rm Im }\Sigma_{{\bf k} + {\bf Q}}|$.
This can be fulfilled for a large  $\xi$ or for a band with
 large effective mass.
Because of the small correlation length of our
calculation, the
shadow states can only be understood from the
above discussed strong deformation
of the quasi particle dispersion.

In  the inset of Fig.~\ref{fig3}, we show the   frequency
dependence of ${\rm Im} \Sigma_{\bf k }(\omega)$
for various  ${\bf k}$-points.
 We find a pronounced difference between   momenta
at the FS  and away from  it.
At the FS, the transition from a low frequency
$\omega^2$ to a linear in $\omega$ behavior
occurs below $0.005 \, {\rm eV}$.
This extremely low energy scale is expected to
be of importance for  the anomalous transport
 properties of the high-T$_c$ materials.
At ${\bf k}=(\pi,\pi/8)$, where the shadow band
crosses the Fermi level, we find a
double well structure.
This reflects the strong coupling of the shadow
states to the main FS and is a precursor
of the singular behavior of ${\rm Im} \Sigma_{\bf k }(\omega)$
in the antiferromagnetic
state~\cite{KS90}.
This anomalous frequency dependence is
different from the main assumption of the
Luttinger theorem~\cite{L61} (${\rm Im}
\Sigma_{\bf k }(\omega) \propto \omega^2$)
and leads
to an occupation of shadow states and consquently
 to a violation of the theorem, shown
in Fig.~\ref{fig1}.
An experimental indication for the violation of
the  Luttinger theorem in underdoped systems
was recently found by Liu {\em et al.}~\cite{LVP92}.

In conclusion, we presented a new numerical
method for the solution of the FLEX-equations
on the real frequency axis, which permits the
analysis of the fine structure of the excitation
spectrum  for high-T$_c$ materials.
A central role for all  results  shown here plays
the exceptional behavior of the quasi
particles at the $X$-point.
The occurrence of a flat dispersion at the
$X$-point was shown to be of extreme importance
for a deep understanding of the shadows of
the Fermi surface without broken antiferromagnetic
order  and
for the violation of the  Luttinger theorem.
All this shows clearly that the quasiparticle
properties of the high-T$_c$ superconductors
are closely intertwined with the strong
antiferromagnetic fluctuations.
In view of the importance of antiferromagnetic
correlations for superconductivity~\cite{D94},
we extended our approach to the superconducting
state, to investigate the phenomena discussed
in this paper below T$_c$.~\cite{GSL95}
%
%
%----References
%
%

 \end{multicols}

\begin{references}
\bibitem[19]{asympt}  Using  FFT, one  assumes the functions to
       be periodic in frequency.  Consequently one has
       to chose the energy interval of the
      calculation       in a way that
      $\Sigma_{\bf k}(\omega)$  is zero at the boundaries,
      fulfilled for
      $\omega \approx \pm 100 t_o$ , since
       it    behaves like $1/\omega$ for large   $|\omega|$.
      Therefore, we calculate the Laplace transform of
      $ \tilde{\Sigma}_{{\bf i}}(t) =\Sigma_{{\bf i}}(t)-
       \Sigma_{{\bf i}}(0)\exp\{-(t/\tau)^2\}$
      with $\tau \approx 2 t_o^{-1}$,  which decreases
      faster than
      $1/\omega$ and the frequency  range can be
      chosen much smaller ($\pm 30 t_o$).
      The  subtracted term  can be transformed analytically.
\bibitem[20]{details} The calculations are performed
                    on a ($64 \times 64$) square lattice.
                    We use 4096 equally spaced energy points
                    in the interval $[-30 t_o,30 t_o]$.
                    In order to gain memory space, we skip most
                    of these points by introducing a logarithmic
                    mesh point scale after FFT, but keep the
                   above point density in the low energy region.
                    In order to perform a later FFT, the skipped
                     points are recalculated by Pad\'e interpolation, leading
                    to a low energy resolution of $0.014 t_o$.
\end{references}
\end{document}